\documentclass[preprint,aps,prb,showpacs,superscriptaddress]{revtex4}
\usepackage{graphicx}
\usepackage{bm}
\begin{document}
\title{Double-walled carbon nanotubes as hundred gigahertz oscillators}
\author{Zhan-chun \surname{Tu}}
\email[Email address: ] {tzc@itp.ac.cn} \affiliation{Institute of
Theoretical Physics,
 The Chinese Academy of Sciences,
 P.O.Box 2735 Beijing 100080, China}
\author{Zhong-can \surname{Ou-Yang}}
\affiliation{Institute of Theoretical Physics,
 The Chinese Academy of Sciences,
 P.O.Box 2735 Beijing 100080, China}
\affiliation{Center for Advanced Study,
 Tsinghua University, Beijing 100084, China}
\begin{abstract}
Based on the van der Waals interaction, the periodically nonlinear
potential of a singe-walled carbon nanotube (SWNT) with finite
length in an infinite length SWNT is analytically obtained. It is
found that the inner SWNT can oscillate in the outer SWNT with
frequency beyond ten Gigahertz, even up to a hundred Gigahertz.
\end{abstract}
 \maketitle
With the discovery of carbon nanotubes (CNT's) in 1991,\cite{s}
the construction of nanodevices will be turned into
reality.\cite{l} Drexler foresaw a double-walled carbon nanotube
(DWNT) as the most efficient bearing for nanomechanical
needs.\cite{Drexler} His suggestion was to take two flat sheets of
graphite, roll them into two cylinders of slightly different
diameter (the diameter difference is about 3.4 \AA), and insert
the smaller one into the larger one. The outer one forms the
sleeve and the inner one forms the shaft. Recently, Cumings and
Zettl \cite{Cumings} reported an important step toward this kind
of nanodevice by constructing the bearings out of multi-walled
carbon nanotubes (MWNT's) with diameters of a few tens of
nanometers. They found that the bearings were the ideal
low-friction and low-wear nanodevices. Zheng and Jiang
\cite{Zheng} proved that the devices of Cumings and Zettl led to
oscillators of frequency up to 1.39 GHz. Intuition tells us that
Drexler's sleeve and shaft should form an oscillator of much
higher frequency. Here, we take a single-walled carbon nanotube
(SWNT) with infinite length as the sleeve and a finite length SWNT
as the shaft. Considering the van der Waals interactions between
the carbon atoms in the sleeve and those in the shaft, we
analytically obtain the periodically nonlinear potential of the
shaft in the sleeve and find that Drexler's sleeve and shaft form
an oscillator of frequency beyond ten Gigahertz, even up to a
hundred Gigahertz.

A SWNT without two caps can be constructed by wrapping up a single
sheet of graphite such that the two equivalent sites of hexagonal
lattice coincide.\cite{r1} The chiral vector ${\bf C}_{h}$, which
defines the relative location of the two cites, is specified by a
pair of integers\cite{ct} $(n, m)$ which is called the index of
SWNT and relates ${\bf C}_{h}$ to the two unit vectors ${\bf
a}_{1}$ and ${\bf a}_{2}$ of graphite (${\bf C}_{h}=n{\bf
a}_{1}+m{\bf a}_{2}$). The translational vector ${\bf T}$
corresponds to the first lattice point of 2D graphitic sheet
through which the line normal to the chiral vector ${\bf C}_{h}$
passes. The unit cell of the SWNT is the rectangle defined by the
vectors ${\bf C}_{h}$ and ${\bf T}$, while the vectors ${\bf
a}_{1}$ and ${\bf a}_{2}$ define the area of the unit cell of 2D
graphite. The number of hexagons per unit cell of SWNT $N$ is
obtained as a function of $n$ and $m$ as $N=2(n^2+m^2+nm)/d_R$,
where $d_R$ is the greatest common divisor of ($2m+n$) and
($2n+m$). There are $2N$ carbon atoms in each unit cell of SWNT
because every hexagon contains two atoms. The symmetry vector
${\bf R}$ is used for generating the coordinates of carbon atoms
in the nanotube and is defined as the site vector having the
smallest component in the direction of ${\bf C}_h$. From a
geometric standpoint, the vector ${\bf R}$ consists of a rotation
around the nanotube axis by an angle $\psi=2\pi/N$ combined with a
translation $\tau$ in the direction of ${\bf T}$; therefore, ${\bf
R}$ can be denoted by ${\bf R}=(\psi|\tau)$. Using the symmetry
vector ${\bf R}$, we can divide $2N$ carbon atoms in the unit cell
of the SWNT into two classes:\cite{tzc2} one includes $N$ atoms
whose site vectors satisfy
\begin{equation}\label{sitea}
{\bf A}_l=l{\bf R}-[l{\bf R}\cdot{\bf T}/{\bf T}^2]{\bf T} \quad
(l=0,1,2,\cdots,N-1),\end{equation} another includes the remain
$N$ atoms whose site vectors satisfy
\begin{eqnarray}\label{siteb} {\bf B}_l&=&l{\bf R}+{\bf
B}_0-[(l{\bf R}+{\bf B}_0)\cdot{\bf T}/{\bf T}^2]{\bf T}\nonumber
\\ &-&[(l{\bf R}+{\bf B}_0)\cdot{\bf C}_h/{\bf C}_h^2]{\bf C}_h
\quad (l=0,1,2,\cdots,N-1),\end{eqnarray}
where {\bf B}$_0$ represents the nearest neighbor atom to {\bf
A}$_0$ in the unit cell. In and only in above two equations,
$[\cdots]$ denotes the Gaussian function, e.g. $[5.3]=5$.

\begin{figure}[htp!]
\includegraphics[width=7cm]{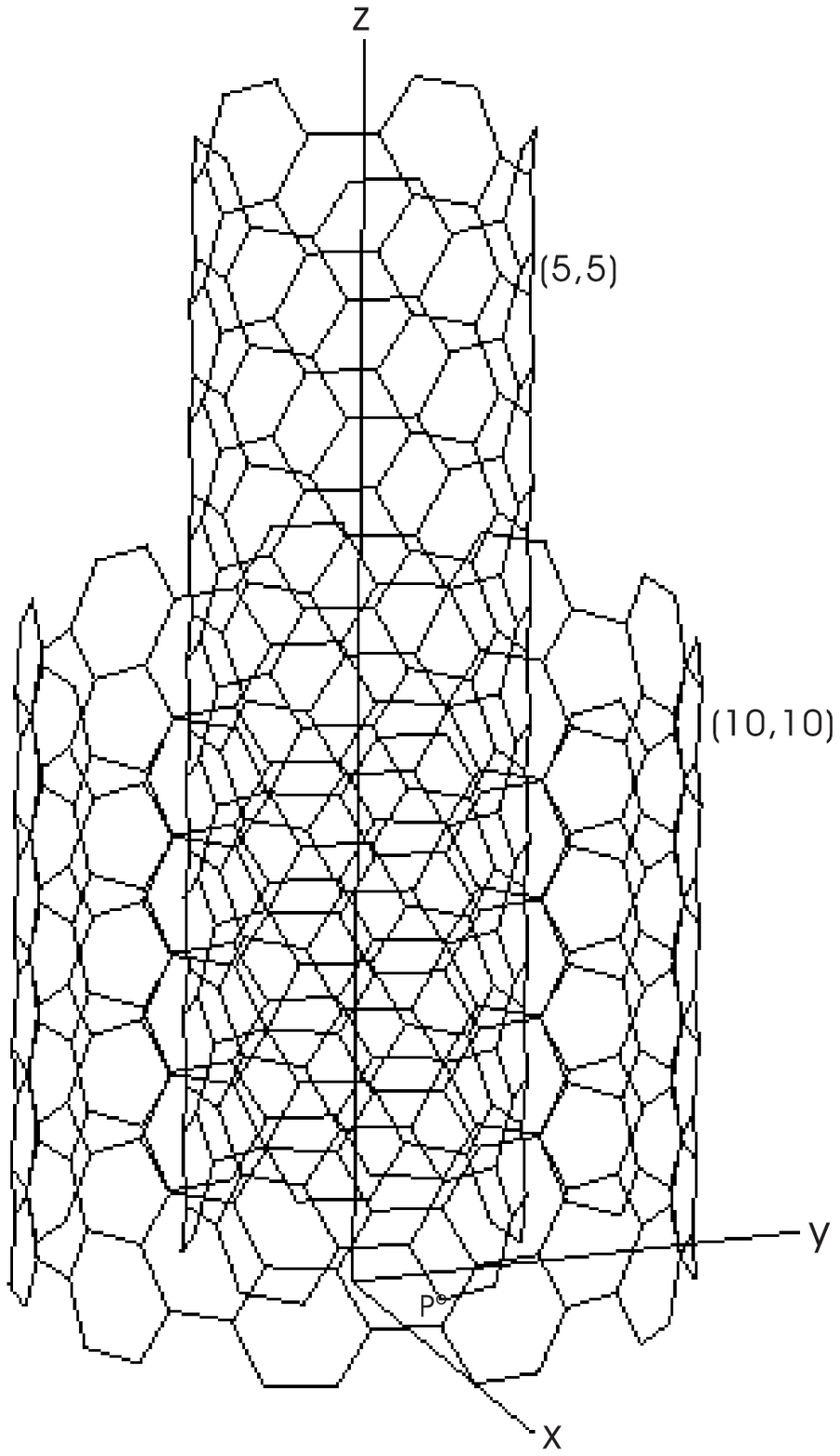} \caption{\label{fig1}Drexler's
sleeve and shaft. The sleeve is the $(10, 10)$ SWNT with infinite
length and the shaft the $(5, 5)$ tube with finite length. The
$z$-axis is the common axis of the $(5, 5)$ and $(10, 10)$ tubes,
and the $x$-axis passes through an atom on the $(10, 10)$ tube. P
is the same type atom on the $(5, 5)$ tube as which the $x$-axis
passes through on the $(10, 10)$ tube.}
\end{figure}

We construct Drexler's sleeve and shaft as FIG.~\ref{fig1}. The
sleeve is the $(10, 10)$ SWNT with infinite length and the shaft
the $(5, 5)$ tube with finite length (e.g. $10T$, where $T=|{\bf
T}|$). The $z$-axis is the common axis of $(5, 5)$ and $(10, 10)$
tubes, and the $x$-axis passes through an atom on the $(10, 10)$
tube. P is the same type atom on the $(5, 5)$ tube as which the
$x$-axis passes through on the $(10, 10)$ tube. The van der Waals
interaction between atoms in the sleeve and the shaft is taken as
Lennard-Jones potential:
\begin{equation} \label{waals}
U(R)=4\epsilon[(\sigma/R)^{12}-(\sigma/R)^6] ,\end{equation} where
R is the distance between atoms in the sleeve and the shaft,
$\epsilon=2.964$ meV and $\sigma=3.4$ \AA.\cite{lu}

\begin{figure}[htp!]
\includegraphics[width=7cm]{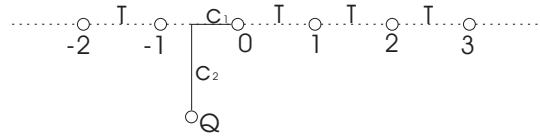}
\caption{\label{fig2}An infinite atom chain and an atom Q out of
the chain. Many atoms distributing regularly in a line form the
infinite atom chain. The interval between the neighbor atoms in
the chain is $T$, and the site of atom Q relative to atom 0 can be
represented by numbers $c_1$ and $c_2$.}
\end{figure}

To obtain the potential of the shaft in the sleeve, we firstly
consider a simplified system shown in FIG.~\ref{fig2}: Many atoms
distributing regularly in a line form an infinite atom chain and
an atom Q is out of the chain. The interval between the neighbor
atoms in the chain is $T$, and the site of atom Q relative to atom
0 can be represented by numbers $c_1$ and $c_2$. If taking the
Lennard-Jones potential between atom Q and atoms in the chain, we
can calculate the potential between atom Q and the chain as
\begin{equation} \label{upc}
U_{QC}=4\epsilon[\sigma^{12}S_6(c_1,c_2)-\sigma^6S_3(c_1,c_2)]
,\end{equation}where
$S_k(c_1,c_2)=\sum_{n=-\infty}^{\infty}\frac{1}{[(c_1+nT)^2+c_2^2]^k}\quad
(k=1,2,\cdots)$ which can be calculated through the following
recursion:\cite{anran}
\begin{equation}\label{sk}
\left\{\begin{array}{l} S_1(c_1,c_2)=\frac{\pi
\sinh (2\pi c_2/T)}{c_2T[\cosh (2\pi c_2/T)-\cos (2\pi c_1/T)]}\\
S_{k+1}(c_1,c_2)=-1/(2kc_2) \partial S_k/\partial c_2 \end{array}
\right. .\end{equation}

\begin{figure}[htp!]
\includegraphics[width=7cm]{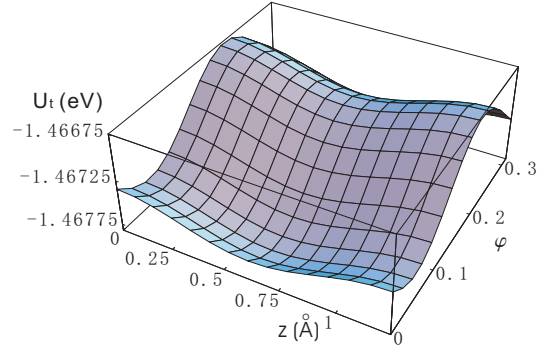}
\caption{\label{fig3}The potential of the shaft in the sleeve. The
potential depends the relative position between the shaft and
sleeve. $z$ and $\varphi$ are the coordinate parameters of atom
P.}
\end{figure}

The sleeve, $(10, 10)$ tube with infinite length can be regarded
as $2No=40$ chains, so the potential of any point Q in the sleeve
can be calculated as $U=\sum_{i=1}^{2No}U_{QC}$. If we denote the
coordinate of atom P in FIG.~\ref{fig1} as $(r\cos\varphi,
r\sin\varphi, z)$ where $r=3.39$ \AA \ \ for $(5, 5)$ tube, the
coordinates of all atoms on the shaft can be calculated from
Eqs.~(\ref{sitea}) and (\ref{siteb}), and their potentials in the
sleeve can be obtained. Summing them, we get the total potential
$U_t(\varphi,z)$ of the shaft in the sleeve which shown in the
FIG.~\ref{fig3} and satisfies
$U_t(\varphi+2\pi/No,z+T/2)=U_t(\varphi,z)$. Here we regard the
sleeve and the shaft as rigid bodies and the sleeve is fixed. The
equilibrium coordinates of the shaft is calculated as
$(\varphi^*,z^*)=(\pi/60, T/4)$ in the range of
$0\leq\varphi<2\pi/No$ and $0\leq z<T/2$. Moreover, from
FIG.~\ref{fig3} we find the shaft just translates along or rotates
around the tube axis if it depart $\delta z=z-z^*$ or
$\delta\varphi=\varphi-\varphi^*$ from the equilibrium point
$(\varphi^*,z^*)$, and it can be regarded as the translational or
rotational oscillator respectively.

\begin{figure}[htp!]
\includegraphics[width=7cm]{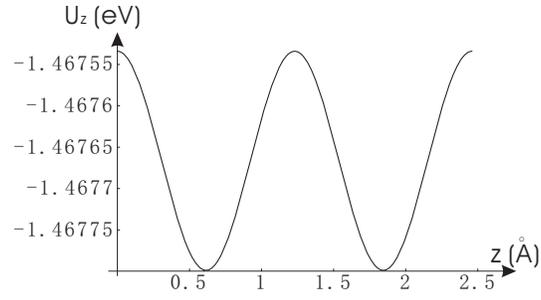}
\caption{\label{fig4}The potential of the shaft in the sleeve
after the departure $\delta z=z-z^*$.}
\end{figure}

\begin{figure}[htp!]
\includegraphics[width=7cm]{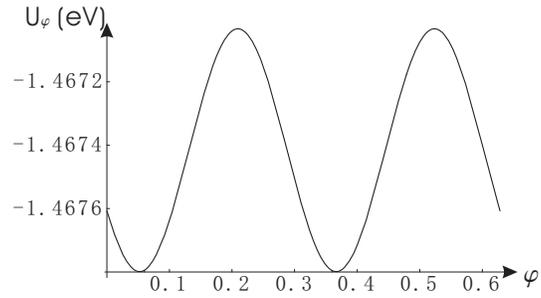}
\caption{\label{fig5}The potential of the shaft in the sleeve
after the departure $\delta\varphi=\varphi-\varphi^*$.}
\end{figure}

\begin{figure}[htp!]
\includegraphics[width=7cm]{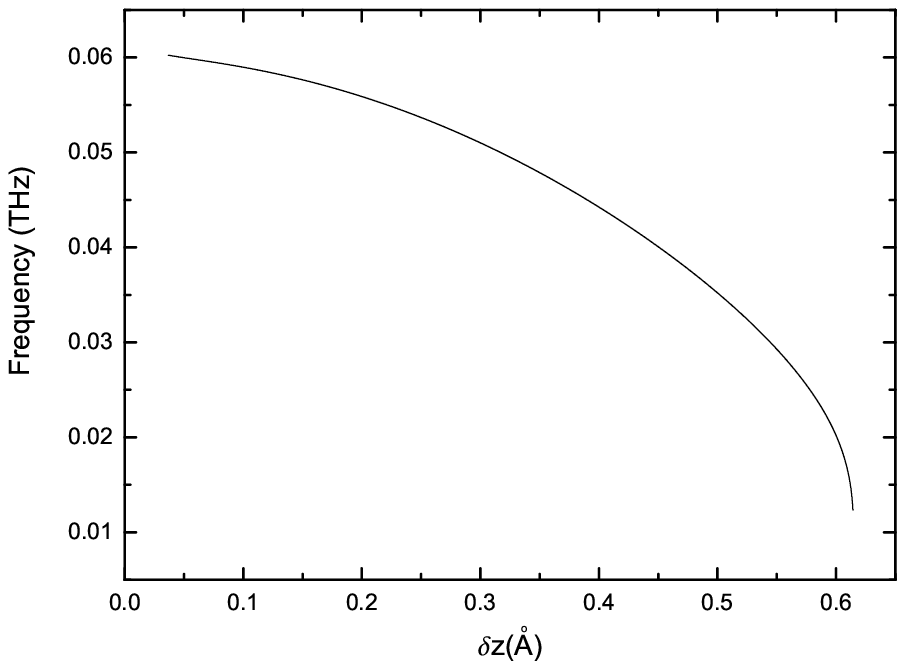}
\caption{\label{fig6}The relation between the frequency of the
translational oscillator and the initial departure along the tube
axis from the equilibrium point.}
\end{figure}

\begin{figure}[htp!]
\includegraphics[width=7cm]{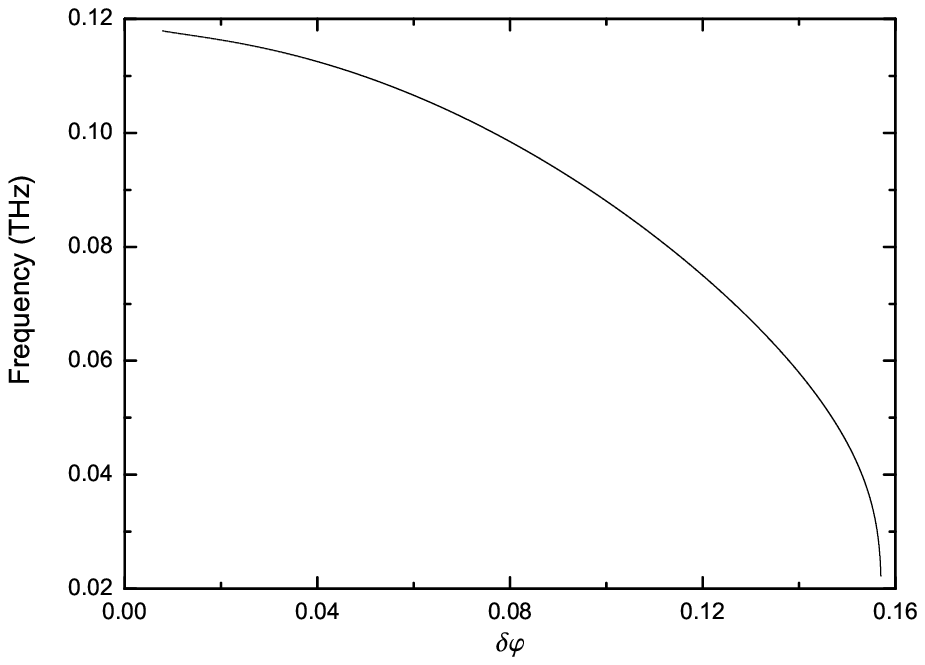}
\caption{\label{fig7}The relation between the frequency of the
rotational oscillator and the initial departure around the tube
axis from the equilibrium point.}
\end{figure}

We plot potentials of the shaft in the sleeve after the departure
in FIG.~\ref{fig4} and FIG.~\ref{fig5}, where
$U_z=U_t(\varphi^*,z)$ and $U_{\varphi}=(\varphi,z^*)$. From $U_z$
and $U_{\varphi}$ we can easily obtain the frequency of the
translational oscillator
\begin{equation}
f_z=1/\{4\int_{z^*}^{z^*+\delta z}\sqrt{\frac{m}{2[U_z(z^*+\delta
z)-U_z(\xi)]}}\ \ d\xi\},\end{equation} as well as the frequency
of the rotational oscillator
\begin{equation}
f_{\varphi}=1/\{4\int_{\varphi^*}^{\varphi^*+\delta
\varphi}\sqrt{\frac{mr^2}{2[U_{\varphi}(\varphi^*+\delta
\varphi)-U_{\varphi}(\xi)]}}\ \ d\xi\},\end{equation} where $m$ is
the mass of the shaft, $0\leq\delta\varphi<\pi/No$ and $0\leq
\delta z<T/4$. The function curves $f_z\sim\delta z$ and
$f_{\varphi}\sim\delta\varphi$ are shown in FIG.~\ref{fig6} and
FIG.~\ref{fig7}, which indicates the oscillator frequencies beyond
ten Gigahertz, even up to a hundred Gigahertz. There are two main
reasons to explain that frequencies we predicted are much higher
than that predicted by Zheng and Jiang:\cite{Zheng} (1) We
consider the DWNT with the outer tube being infinite length, but
Zheng and Jiang consider the finite length MWNT with many layers.
(2) Zheng and Jiang use the linear potential between the core and
the outer MWNT's, but our calculation gives the periodically
nonlinear potential between the inner and the outer SWNT's.
Fortunately, we find the latest paper of Legoas \textit{et
al.}\cite{Legoas} when we finish this paper. In the abstract of
their paper, Legoas \textit{et al.} alleged their observation of
frequencies as large as 38 GHz through the molecular-dynamics
simulations, which agrees with our results. But in the body of
their paper, they might forget to mention this matter.

In summary, considering the van der Waals interactions between the
carbon atoms in the sleeve and those in the shaft, we analytically
obtain the periodically nonlinear potential of the shaft in the
sleeve and find that Drexler's sleeve and shaft form an oscillator
of frequency beyond ten Gigahertz, even up to a hundred Gigahertz.
In above discussion, we do not consider the friction because it is
much smaller than the van der Waals force.\cite{Cumings} But we
think it is worth studying the following questions: What is the
essence of the friction in the mesoscopic systems? What is its
dissipative mechanism? Is there any relation between it and the
van der Waals force? Why is it so small? Moreover, the procedure
to analytically obtain the periodically nonlinear potential will
be useful to seek the behavior of gas or water molecules
\cite{Liu,Gordillo,Stan,Zhao,Koga,Hummer} in the SWNT.

The authors acknowledge Miss R. An's help in mathematics.

\newpage

\end{document}